\documentclass[preprint,superscriptaddress,showpacs,preprintnumbers,amsmath,amssymb]{revtex4}

\usepackage{graphicx}
\usepackage{dcolumn}
\usepackage{bm}

\def\lesssim{\ \raise.3ex\hbox{$<$}\kern-0.8em\lower.7ex\hbox{$\sim$}\ }
\def\gesim{\ \raise.3ex\hbox{$>$}\kern-0.8em\lower.7ex\hbox{$\sim$}\ }

\def\rnum#1{\expandafter{\romannumeral #1}} 
\def\Rnum#1{\uppercase\expandafter{\romannumeral #1}} 

\begin{document}

\title{Superfluid density of states and pseudogap phenomenon in the BCS-BEC crossover regime of a superfluid Fermi gas}
\author{Ryota Watanabe}
\affiliation{Faculty of Science and Technology, Keio University,
3-14-1 Hiyoshi, Kohoku-ku, Yokohama 223-8522, Japan}
\email{rwatanab@mail.rk.phys.keio.ac.jp}

\author{Shunji Tsuchiya}
\affiliation{Department of Physics, Faculty of Science, Tokyo University
of Science, 1-3 Kagurazaka, Shinjuku-ku, Tokyo 162-8601, Japan}
\affiliation{CREST(JST), 4-1-8 Honcho, Saitama 332-0012, Japan}

\author{Yoji Ohashi}
\affiliation{Faculty of Science and Technology, Keio University,
3-14-1 Hiyoshi, Kohoku-ku, Yokohama 223-8522, Japan}
\affiliation{CREST(JST), 4-1-8 Honcho, Saitama 332-0012, Japan}

\date{\today}       
\begin{abstract}
We investigate single-particle excitations and strong-coupling effects in the BCS-BEC crossover regime of a superfluid Fermi gas. Including phase and amplitude fluctuations of the superfluid order parameter within a $T$-matrix theory, we calculate the superfluid density of states (DOS), as well as single-particle spectral weight, over the entire BCS-BEC crossover region below the superfluid transition temperature $T_{\rm c}$. We clarify how the pseudogap in the normal state evolves into the superfluid gap, as one passes through $T_{\rm c}$. While the pseudogap in DOS continuously evolves into the superfluid gap in the weak-coupling BCS regime, the superfluid gap in the crossover region is shown to appear in DOS after the pseudogap disappears below $T_{\rm c}$. In the phase diagram with respect to the temperature and interaction strength, we determine the region where strong pairing fluctuations dominate over single-particle properties of the system. Our results would be useful for the study of strong-coupling phenomena in the BCS-BEC crossover regime of a superfluid Fermi gas. 
\end{abstract}
\maketitle

\section{Introduction}
\label{section1}
Strong correlation between particles is one of the most important key
issues in condensed matter physics. The recently developed ultracold
Fermi gases offer unique opportunities to study this important topic in
a controlled manner, by maximally using highly tunable physical
parameters. Indeed, the crossover from the Bardeen-Cooper-Schrieffer
(BCS) type superfluid to the Bose-Einstein condensation (BEC) of tightly
bound molecules\cite{Eagles,Leggett,NSR,SadeMelo} has been realized in
$^{40}$K\cite{Regal} and $^6$Li Fermi
gases\cite{Zwierlein,Kinast,Bartenstein}, using a tunable pairing
interaction associated with a Feshbach
resonance\cite{Timmermans,Holland,Ohashi,Giorgini,Bloch,Ketterle}. This
BCS-BEC crossover demonstrates the usefulness of an ultracold Fermi gas
as a quantum simulator for strongly correlated fermion systems. In
particular, this system is expected to be useful for the study of
high-$T_{\rm c}$ cuprates\cite{Lee}.
\par
The recent momentum-resolved photoemission-type spectroscopy developed by JILA group\cite{Stewart} is a powerful technique to probe microscopic properties of a cold Fermi gas in the crossover region\cite{Chen,Tsuchiya2}. This experiment is an analogue of the angle-resolved photoemission spectroscopy (ARPES) in condensed matter physics\cite{Damascelli}. In the photoemission-type experiment developed by JILA group\cite{Stewart}, atoms are transferred to the third empty atomic state by rf-pulse. Using this, one can directly measure the single-particle spectral weight (SW), as well as the occupied density of states (DOS). As a remarkable experimental result, the pseudogap has been observed in $^{40}$K Fermi gas\cite{Gaebler}. The back-bending curve of single-particle dispersion has been observed as a characteristic signature of pseudogap phenomenon\cite{Gaebler,Tsuchiya2}. 
\par
The pseudogap has been considered as a crucial key issue in the
underdoped regime of high-$T_{\rm c}$ cuprates\cite{Lee,Fischer,Damascelli}. So far, various mechanisms have been proposed to explain this phenomenon, such as preformed Cooper pairs\cite{Pines,Kampf,Randeria,Singer,Janko,Rohe,Yanase,Yanase2,Perali}, antiferromagnetic spin fluctuations\cite{Pines}, localization of Cooper pairs\cite{Yanase3,Fischer}, and hidden order\cite{Chakravarty}. However, the complete understanding has not been obtained yet, because of the complexity of high-$T_{\rm c}$ cuprates. In contrast, the origin of the pseudogap observed in a $^{40}$K Fermi gas\cite{Stewart,Gaebler} is well understood. Namely, preformed pairs associated with strong pairing fluctuations are responsible for this phenomenon\cite{Tsuchiya}. Thus, cold Fermi gases are very suitable for the study of preformed pair scenario proposed in high-$T_{\rm c}$ cuprates. 
\par
Recently, the pseudogap phenomenon above $T_{\rm c}$ has been theoretically addressed in the literature of cold Fermi gas\cite{Chen,Tsuchiya,Haussmann,Tsuchiya2,Hu,Chien,Su}. It has been shown that a gap like structure emerges in the single-particle excitation spectra in the pseudogap regime\cite{Tsuchiya,Chien,Su}. It has been also pointed out the existence of two pseudogap temperatures $T^*$ and $T^{**}$\cite{Tsuchiya}: While a dip structure appears in DOS below $T^\ast$, a double-peak structure and back-bending dispersion are seen in SW below $T^{\ast\ast}$. 
\par
In this paper, we investigate single-particle excitations and effects of strong pairing fluctuations in the BCS-BEC crossover regime of a superfluid Fermi gas. Extending our previous paper for the pseudogap phenomenon above $T_{\rm c}$\cite{Tsuchiya} to the superfluid phase below $T_{\rm c}$, we calculate DOS within a $T$-matrix theory. We clarify how the pseudogap above $T_{\rm c}$ evolves into the superfluid gap below $T_{\rm c}$. While the evolution is continuous in the weak-coupling BCS regime, the superfluid gap is shown to appear after the pseudogap disappears below $T_{\rm c}$ in the crossover region. We also identify the region where pairing fluctuations dominate over single-particle properties in the phase diagram with respect to the temperature and interaction strength. Recently, strong-coupling effects on SW has been discussed below $T_{\rm c}$\cite{Pieri}. In this paper, we also treat this quantity to examine how the pseudogap in DOS is related to SW affected by pairing fluctuations. 
\par
This paper is organized as follows. In Sec.~\ref{section2}, we explain our formulation based on a $T$-matrix theory. In Sec.~\ref{section3}, we present our numerical results for the superfluid DOS, as well as SW, to discuss strong-coupling effects on these quantities. In Sec.~\ref{section4}, we present the phase diagram of a superfluid Fermi gas to clarify the region where pairing fluctuations are crucial for single-particle excitations. Throughout this paper, we set $\hbar=k_B=1$, and the system volume $V$ is taken to be unity.

\section{Formalism}
\label{section2}
We consider a three-dimensional Fermi gas, consisting of two atomic hyperfine states described by pseudospin $\sigma=\uparrow,\downarrow$. We assume that the two hyperfine states are equally populated, and ignore effects of a harmonic trap, for simplicity. We also assume a broad Feshbach resonance as the origin of tunable pairing interaction. In this case, it is known that the detailed Feshbach mechanism is not crucial for the study of interesting BCS-BEC crossover physics, so that we can safely use the ordinary single-channel BCS model, given by
\begin{equation}
H=\sum_{{\bm p},\sigma}\xi_p c_{\bm p\sigma}^\dagger c_{\bm p\sigma}
-U\sum_{\bm p,\bm p',\bm q}c_{{\bm p}+{\bm q}/2\uparrow}^\dagger
 c_{-{\bm p}+{\bm q}/2\downarrow}^\dagger c_{-{\bm p}'+{\bm
 q}/2\downarrow}c_{{\bm p}'+{\bm q}/2\uparrow}.
\label{ham0}
\end{equation}
Here, $c^\dagger_{\bm p\sigma}$ is the creation operator of a Fermi atom with pseudospin $\sigma=\uparrow,\downarrow$ and the kinetic energy $\xi_p=\epsilon_p-\mu=\frac{p^2}{2m}-\mu$ measured from the chemical potential $\mu$ (where $m$ is an atomic mass). $-U(<0)$ is a tunable pairing interaction associated with a Feshbach resonance. In cold atom physics, this pairing interaction is conveniently measured in terms of the $s$-wave scattering length $a_s$, which is related to $U$ as\cite{Randeria2},
\begin{equation}
\frac{4\pi a_s}{m}=-\frac{U}{1-U\sum_{\bm
p}^{\omega_c}\frac{1}{2\epsilon_p}},
\label{as}
\end{equation}
where $\omega_c$ is a high-energy cutoff. In this scale, the weak-coupling BCS regime and the strong-coupling BEC regime are, respectively, given by $(k_{\rm F}a_s)^{-1}\lesssim -1$ and $(k_{\rm F}a_s)^{-1}\gesim1$, where $k_{\rm F}$ is the Fermi momentum. The region $-1\lesssim (k_{\rm F}a_s)^{-1}\lesssim 1$ is called the crossover region.
\par
To consider fluctuations in the Cooper channel below $T_{\rm c}$, it is convenient to rewrite Eq. (\ref{ham0}) into the form consisting of the mean-field part and fluctuation contribution. Introducing the Nambu field, 
\begin{equation}
\Psi_{\bm p}=
\left(
\begin{array}{c}
c_{{\bm p}\uparrow}\\
c_{-{\bm p}\downarrow}^\dagger
\end{array}
\right),
\label{nambu}
\end{equation}
we have\cite{Takada} 
\begin{equation}
H=\sum_{\bm p}\Psi_{\bm p}^\dagger[\xi_p\tau_3-\Delta\tau_1]\Psi_{\bm
p}-U\sum_{{\bm q}}\rho_+({\bm q})\rho_-(-{\bm q}).
\label{ham}
\end{equation}
Here, $\tau_j$ ($j=1,2,3$) are Pauli matrices acting on particle-hole space. $\rho_\pm({\bm q})\equiv[\rho_1({\bm q})\pm i\rho_2({\bm q})]/2$ are the generalized density operators, where $\rho_j({\bm q})=\sum_{\bm p}\Psi^\dagger_{{\bm p}+{\bm q}/2}\tau_j\Psi_{{\bm p}-{\bm q}/2}$. In Eq. (\ref{ham}), the first term is the mean-field Hamiltonian, where the superfluid order parameter $\Delta\equiv U\sum_{\bm p}\langle c_{-{\bm p}\downarrow} c_{{\bm p}\uparrow} \rangle$ is taken to be real and proportional to the $\tau_1$-component. In this choice, $\rho_1({\bm q})$ and $\rho_2({\bm q})$ physically describe the amplitude and phase fluctuations of the order parameter, respectively\cite{Takada,noteZ}. Namely, the last term in Eq. (\ref{ham}) describes effects of pairing fluctuations. 
\par
\begin{figure}[t]
\includegraphics[width=10cm]{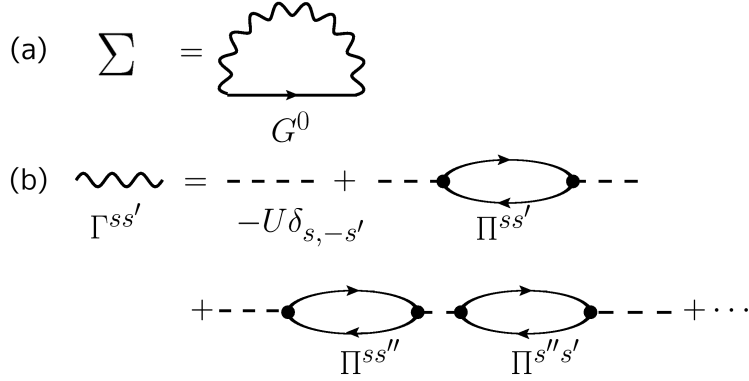}
\caption{
Fluctuation contributions to the self-energy $\Sigma_{\bm p}(i\omega_n)$ in the $T$-matrix approximation employed in this paper. (a) Self-energy correction. (b) Particle-particle scattering matrix $\Gamma_{\bm q}(i\nu_n)$. The solid and dashed lines represent the mean-field Green's function $G_{\bm p}^0(i\omega_n)$ and pairing interaction $-U$, respectively. The bubble diagrams represent the zeroth order correlation functions $\Pi_{\bm q}^{ss'}(i\nu_n)$ (where $s=\pm$), describing pairing fluctuations below $T_{\rm c}$. The solid circles are Pauli matrices $\tau_s$.}
\label{fig1}
\end{figure}

In this paper, we take into account the last term in Eq.~(\ref{ham}) within the $T$-matrix approximation\cite{Pieri}. For this purpose, we introduce the $2\times2$-matrix single-particle thermal Green's function, given by
\begin{equation}
G_{\bm p}(i\omega_n)=\frac{1}{G^0_{\bm p}(i\omega_n)^{-1}-\Sigma_{\bm p}(i\omega_n)}.
\label{Green}
\end{equation}
Here, $G^0_{\bm p}(i\omega_n)^{-1}\equiv i\omega_n-\xi_p\tau_3+\Delta\tau_1$ is the mean-field Green's function, where $\omega_n$ is the fermion Matsubara frequency. The $2\times2$-matrix self-energy $\Sigma_{\bm p}(i\omega_n)$ describes fluctuation corrections. Within the $T$-matrix theory, it is diagrammatically given by Fig.~\ref{fig1}\cite{Pieri}. Summing up the diagrams in Fig.~\ref{fig1}, we obtain
\begin{eqnarray}
\Sigma_{\bm p}(i\omega_n)=-
{1 \over \beta}\sum_{\bm q,\nu_n}
\sum_{s,s'=\pm}\Gamma_{\bm q}^{s s'}(i\nu_n)
\tau_{-s}G_{\bm p+\bm q}^0(i\omega_n+i\nu_n)\tau_{-s'},
\label{self-energy}
\end{eqnarray}
where $\beta=1/T$ is the inverse temperature. $\nu_n$ is the boson Matsubara frequency, and $\tau_\pm=[\tau_1\pm i \tau_2]/2$. The particle-particle scattering matrix $\Gamma_{\bm q}^{ss'}(i\nu_n)$ is given by
\begin{eqnarray}
\left(
\begin{array}{cc}
\Gamma_{\bm q}^{+-}(i\nu_n)&
\Gamma_{\bm q}^{++}(i\nu_n)\\
\Gamma_{\bm q}^{--}(i\nu_n)&
\Gamma_{\bm q}^{-+}(i\nu_n)\\
\end{array}
\right)=
-U
\left[
1+U
\left(
\begin{array}{cc}
\Pi_{\bm q}^{+-}(i\nu_n)&
\Pi_{\bm q}^{++}(i\nu_n)\\
\Pi_{\bm q}^{--}(i\nu_n)&
\Pi_{\bm q}^{-+}(i\nu_n)\\
\end{array}
\right)
\right]^{-1}.
\label{gamma}
\end{eqnarray}
Here, $\Pi_{\bm q}^{ss'}(i\nu_n)$ is the lowest-order of the following correlation function:
\begin{equation}
\Pi_{\bm q}^{ss'}(i\nu_n)
=\int_0^\beta d\tau
\langle
T_{\tau}
\{
\rho_s({\bm q},\tau)\rho_{s'}({\bm q},0)
\}
\rangle
e^{i\nu_n\tau}.
\label{corr}
\end{equation}
Evaluating Eq. (\ref{corr}) within the zeroth order with respect to the last term in Eq. (\ref{ham}), we have
\begin{equation}
\Pi_{\bm q}^{ss'}(i\nu_n)={1 \over \beta}\sum_{\bm p,\omega_n}{\mathrm Tr}
\Bigl[
\tau_s G_{\bm p+\bm q/2}^0(i\omega_n+i\nu_n)
\tau_{s'} G_{\bm p-\bm q/2}^0(i\omega_n)\Bigr]. 
\label{polari}
\end{equation}
Executing the $\omega_n$-summation in Eq. (\ref{polari}), we obtain

\begin{eqnarray}
\Pi_{\bm q}^{++}(i\nu_n)&=&\frac{1}{4}\sum_{s=\pm1}\sum_{\bm
 p}\frac{s\Delta^2}{E_{\bm p+\bm q/2}E_{\bm p-\bm q/2}}\frac{E_{\bm
 p+\bm q/2}+sE_{\bm p-\bm q/2}}{\nu_n^2+(E_{\bm p+\bm q/2}+sE_{\bm
 p-\bm q/2})^2}
\nonumber\\ 
&&\qquad\times\left[\tanh\left({{\beta \over 2}E_{\bm p+\bm q/2}}\right)+s\tanh\left({{\beta \over 2}E_{\bm p-\bm q/2}}\right)\right],
\label{explicitpi1}
\end{eqnarray}
\begin{eqnarray}
\Pi_{\bm q}^{+-}(i\nu_n)&=&\frac{1}{4}\sum_{s=\pm1}\sum_{\bm
 p}\left[\left(1+s\frac{\xi_{\bm p+\bm q/2}\xi_{\bm
		     p-\bm q/2}}{E_{\bm p+\bm q/2}E_{\bm
		     p-\bm q/2}}\right)\frac{1}{i\nu_n-(E_{\bm
 p+\bm q/2}+sE_{\bm p-\bm q/2})}\right.\nonumber\\ 
&&+\left.\left(1-\frac{\xi_{\bm p+\bm q/2}}{E_{\bm p+\bm
	  q/2}}\right)\left(1-s\frac{\xi_{\bm p-\bm q/2}}{E_{\bm
	  p-\bm q/2}}\right)\frac{i\nu_n}{\nu_n^2+(E_{\bm p+\bm q/2}+sE_{\bm
	  p-\bm q/2})^2} \right]\nonumber\\ 
&&\qquad\times\left[\tanh\left({\beta \over 2}E_{\bm p+\bm q/2}\right)+s\tanh\left({\beta \over 2}E_{\bm p-\bm q/2}\right)\right],
\label{explicitpi2}
\end{eqnarray}
where $E_{\bm p}=\sqrt{\xi_{\bm p}^2+\Delta^2}$ is the Bogoliubov single-particle excitation spectrum. The other components are given by $\Pi_{\bm q}^{--}(i\nu_n)=\Pi_{\bm q}^{++}(i\nu_n)$, and $\Pi_{\bm q}^{-+}(i\nu_n)=\Pi_{\bm q}^{+-}(-i\nu_n)$.
\par

\begin{figure}
\includegraphics[width=9cm]{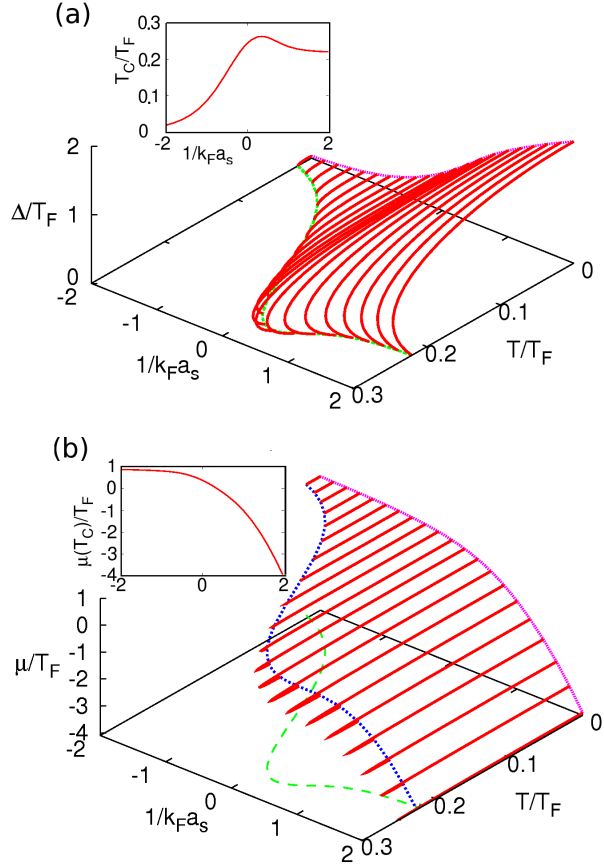}
\caption{(Color online) Calculated superfluid order parameter $\Delta$ (a) and Fermi chemical potential $\mu$ (b) in the BCS-BEC crossover normalized by the Fermi temperature $T_{\rm F}$. We use these results in calculating DOS in Sec. III. The upper and lower insets show $T_{\rm c}$ and $\mu(T_{\rm c})$, respectively. ($T_{\rm c}$ is also shown in panel (b) as the dashed line.) The first-order behavior seen in the crossover region ($(k_Fa_s)^{-1}\gtrsim -0.5$) is an artifact of the $T$-matrix approximation we are using in this paper.
}
\label{fig2}
\end{figure}

\par
DOS is obtained from the analytic continued Green's function, as
\begin{equation}
\rho(\omega)=\sum_{\bm p}A({\bm p},\omega),
\label{DOSg}
\end{equation}
where SW $A({\bm p},\omega)$ has the form
\begin{equation}
A({\bm p},\omega)=-\frac{1}{\pi}\mathrm{Im}G_{\bm p}(i\omega_n\to\omega_+=\omega+i\delta)|_{11}.
\label{SWg}
\end{equation}
The analytic continued self-energy $\Sigma_{\bm p}(\omega\to\omega_+)$ involved in $G_{\bm p}(i\omega\to\omega_+)$ is given by
\begin{eqnarray}
\label{ana0}
\Sigma_{\bm p}(i\omega_n\to \omega_+)|_{11}&=&\Sigma_{\rm HF}\nonumber\\
&+&{1 \over \pi}
\sum_{{\bm q},s=\pm 1}
\int_{-\infty}^\infty
dz
{n_B(z)+f(sE_{{\bm p}+{\bm q}}) \over z-sE_{{\bm p}+{\bm q}}+\omega_+}\left(1-s{\xi_{{\bm p}+{\bm q}} \over E_{{\bm p}+{\bm q}}}\right){\rm Im}\left[\Gamma_{\bm q}^{+-}(z_+)\right],\\
\Sigma_{\bm p}(i\omega_n\to \omega_+)|_{12}&=&
{1 \over \pi}
\sum_{{\bm q},s=\pm 1}
\int_{-\infty}^\infty
dz
{n_B(z)+f(sE_{{\bm p}+{\bm q}}) \over z-sE_{{\bm p}+{\bm q}}+\omega_+}s{\Delta \over E_{{\bm p}+{\bm q}}}{\rm Im}\left[\Gamma_{\bm q}^{++}(z_+)\right],
\label{ana1}
\end{eqnarray}
where $z_+=z+i\delta$, and 
\begin{equation}
\Sigma_{\rm HF}=-{U \over 2}\tau_3\sum_{\bm p}
\left[
1-{\xi_{\bm p} \over E_{\bm p}}\tanh
\left({\beta E_{\bm p} \over 2}\right)
\right]
\label{HF}
\end{equation}
is the Hartree self-energy.
The other components are given by $\Sigma_{\bm p}(\omega_+)|_{22}=-\Sigma_{\bm p}(-\omega_+)|_{11}$, and $\Sigma_{\bm p}(\omega_+)|_{21}=\Sigma_{\bm p}(\omega_+)|_{12}$.
In Eqs. (\ref{ana0}) and (\ref{ana1}), we have carried out the Matsubara frequency summation by using the spectral representation of $\Gamma_{\bm q}^{ss'}(i\nu_q)$, given by
\begin{eqnarray}
\begin{array}{l}
\displaystyle
\Gamma_{\bm q}^{+-}(i\nu_n)
=-U
-{1 \over \pi}
\int_{-\infty}^\infty dz
{
{\rm Im}
\left[\Gamma_{\bm q}^{+-}(i\nu_n\to z_+)
\right]
\over
i\nu_n-z
},
\\
\displaystyle
\Gamma_{\bm q}^{++}(i\nu_n)
=
-{1 \over \pi}
\int_{-\infty}^\infty dz
{
{\rm Im}
\left[\Gamma_{\bm q}^{++}(i\nu_n\to z_+)
\right]
\over
i\nu_n-z
}.
\end{array}
\label{eq1}
\end{eqnarray}
\par
We actually calculate Eq. (\ref{DOSg}) after determining the superfluid order parameter $\Delta$ and Fermi chemical potential $\mu$ below $T_{\rm c}$. In the present $T$-matrix theory, they are obtained by solving the gap equation, 
\begin{equation}
1=U\sum_{\bm p}{1 \over 2E_p}\tanh\frac{E_p}{2T},
\label{gapeq}
\end{equation}
together with the equation for the number of Fermi atoms,
\begin{equation}
N={2 \over \beta}\sum_{\bm p,\omega_n}G_{\bm p}(i\omega_n)|_{11}
e^{i\delta\omega_n}.
\label{number}
\end{equation}
This framework is a natural extension of the Gaussian fluctuation theory developed by Nozi\`{e}res and Schmitt-Rink (NSR)\cite{NSR}, where the self-energy correction $\Sigma_{\bm p}(i\omega_n)$ is taken into account up to the first order. This $T$-matrix theory can properly describe the BCS-BEC crossover behaviors of $T_{\rm c}$ and $\mu$\cite{Perali,Tsuchiya}. (See the insets in Fig.~\ref{fig2}.) In addition, this theory is consistent with the Goldstone's theorem, in the sense that the particle-particle scattering matrix $\Gamma_{\bm q}^{\pm\pm}(i\nu_n)$ in Eq. (\ref{gamma}) has a pole at ${\bm q}=\nu_n=0$. Indeed, the condition that Eq. (\ref{gamma}) has a pole at ${\bm q}=\nu_n=0$ gives
\begin{equation}
\left[
1+U[\Pi_{{\bm q}=0}^{++}(0)+\Pi_{{\bm q}=0}^{+-}(0)]
\right]
\left[
1-U[\Pi_{{\bm q}=0}^{++}(0)-\Pi_{{\bm q}=0}^{+-}(0)]
\right]=0.
\label{check1}
\end{equation}
One finds from Eqs.~(\ref{explicitpi1}) and (\ref{explicitpi2}) that the factor $1-U[\Pi_{{\bm q}=0}^{++}(0)-\Pi_{{\bm q}=0}^{+-}(0)]$ vanishes identically when the gap equation (\ref{gapeq}) is satisfied. We also note that, as in the NSR theory\cite{Fukushima}, the present $T$-matrix theory also shows the first-order phase transition in the BCS-BEC crossover region (See Fig.~\ref{fig2}.), which is, however, an artifact of the theory. To overcome this problem, one needs to include many-body scattering effect between molecules in a consistent manner\cite{OhashiZ}. Although this is an important problem in the BCS-BEC crossover physics, in this paper, we leave it as a future problem and simply use $\Delta$ and $\mu$ in Fig.~\ref{fig2} to examine strong-coupling effects on single-particle excitations below $T_{\rm c}$.

\begin{figure}
\includegraphics[width=\linewidth]{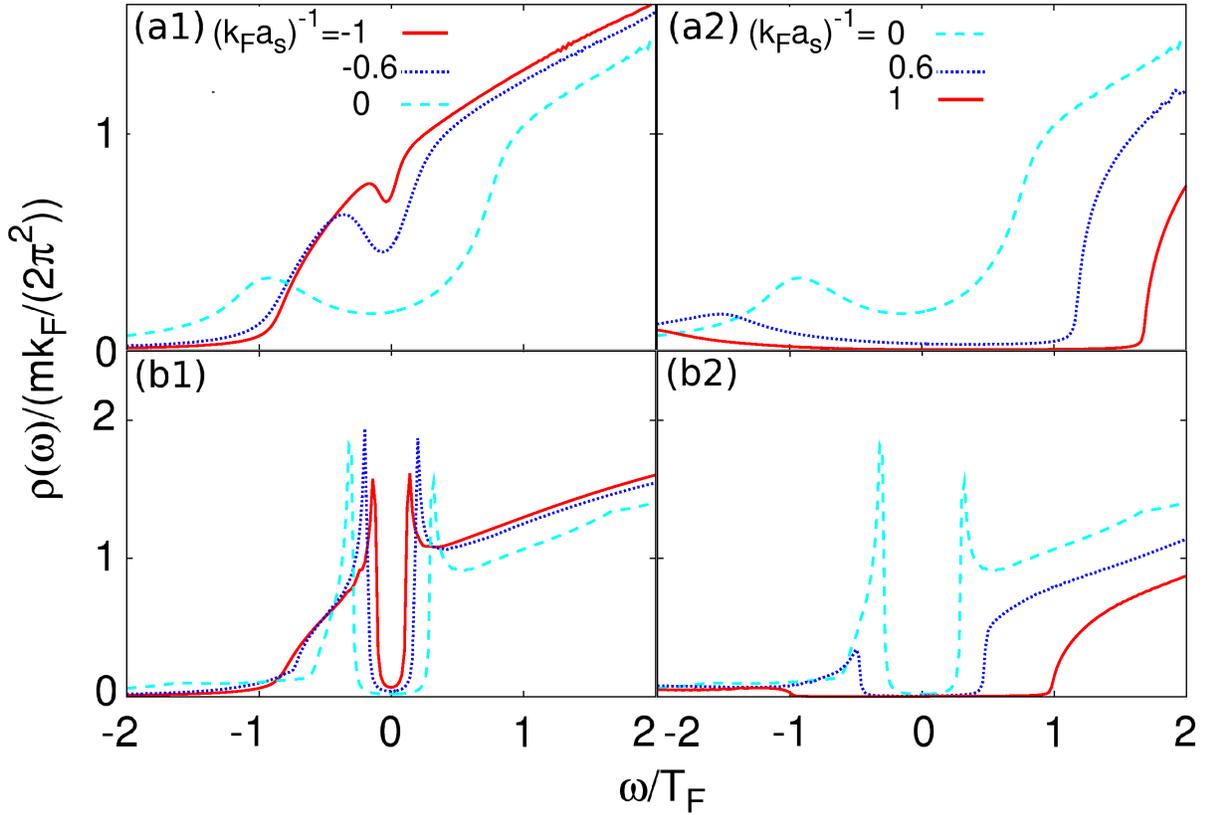}
\caption{(Color online) Upper panels: Pseudogap in DOS $\rho(\omega)$ at $T_{\rm c}$. Lower panels: DOS at $T=0$. In this figure, the left and right panels show the results in the BCS side ($(k_Fa_s)^{-1}\leq 0$) and in the BEC side ($(k_Fa_s)^{-1}\geq 0$), respectively. At $T=0$, small but finite intensity at $\omega\sim 0$ seen in panel (b1) is due to a small imaginary part ($\delta=0.01T_{\rm F}$) introduced to the energy in numerical calculations.}
\label{fig3}
\end{figure}

\section{Single-particle properties in the superfluid phase}
\label{section3}

Figure \ref{fig3} shows DOS at $T_{\rm c}$ (upper panels), as well as at $T=0$ (lower panels), in the BCS-BEC crossover. In the upper panels, the pseudogap structure can be seen around $\omega=0$\cite{Tsuchiya}. This structure becomes more remarkable with increasing the interaction strength, reflecting the enhancement of pairing fluctuations. On the other hand, since thermal fluctuations are absent at $T=0$, the well-known superfluid excitation gap associated with the superfluid order parameter $\Delta$ appears in the lower panels in Fig.~\ref{fig3}. The goal of this section is to show how the pseudogap at $T_{\rm c}$ evolves into the superfluid gap below $T_{\rm c}$.
\par
\begin{figure}
\includegraphics[width=10.0cm]{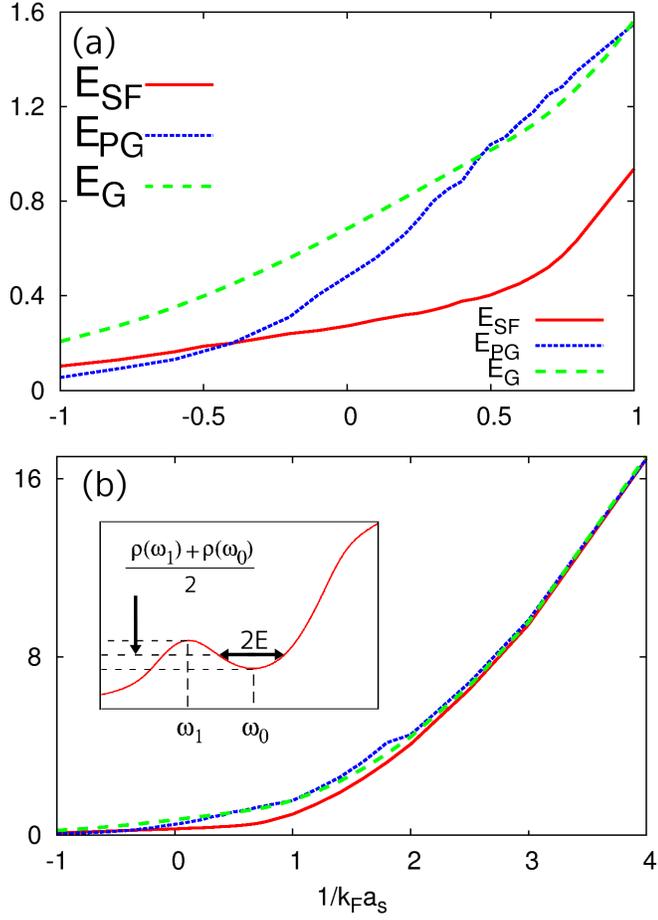}
\caption{(Color online) (a) Comparison of the pseudogap size $E_{\rm PG}$ at $T_{\rm c}$ and superfluid gap size $E_{\rm SF}$ at $T=0$ evaluated from Fig.~\ref{fig3}. For comparison, we also show the energy gap $E_G$ in the BCS-Leggett crossover theory\cite{Leggett} (which equals $\Delta$ when $\mu>0$, and equals $\sqrt{\mu^2+\Delta^2}$ when $\mu<0$). The panel (b) shows the behaviors of these quantities in the strong-coupling BEC regime, where one finds $E_{\rm PG}\simeq E_{\rm SF}\simeq E_G$ when $(k_{\rm F}a_s)^{-1}\gesim 3$. The inset shows how to determine $E_{\rm PG}$ and $E_{\rm SF}$ from Fig.~\ref{fig3}. Since the pseudogap actually does not have a clear energy gap, we conveniently define the gap size as the half of the dip size at $[\rho(\omega_0)+\rho(\omega_1)]/2$, as shown in the inset, where $\omega_0$ and $\omega_1$ are the bottom energy and lower peak position, respectively.
}
\label{fig4}
\end{figure}

In considering this problem, we first note the following two key issues which can be seen in Fig.~\ref{fig3}. The first one is that the size difference between the pseudogap $E_{\rm PG}$ at $T_{\rm c}$ and the superfluid gap $E_{\rm SF}$ at $T=0$ evaluated from Fig.~\ref{fig3} strongly depends on the interaction strength. As shown in Fig.~\ref{fig4}(a), while $E_{\rm PG}$ is smaller than $E_{\rm SF}$ in the weak-coupling BCS regime, the former becomes larger than the latter in the BCS-BEC crossover region. This implies that, while the pseudogap in the BCS regime may smoothly change into the superfluid gap below $T_{\rm c}$, the large pseudogap at $T_{\rm c}$ in the crossover region needs to shrink below $T_{\rm c}$, in addition to the opening of the superfluid gap. 
\par

The second key issue in Fig.~\ref{fig3} is that, although the pseudogap structure in DOS looks similar to the superfluid gap, the former does not have the coherence peaks at the gap edges. However, even in the superfluid phase below $T_{\rm c}$, the coherence peaks are known to disappear by strong-coupling effects. In the present case, these strong-coupling effects involve pairing fluctuations excited thermally and the formation of tightly bound molecules. While the former effects are expected only at finite temperatures, the latter may exist down to $T=0$ in the BEC regime. The latter effect can be easily confirmed by using the superfluid DOS within the mean-field theory, given by 
\begin{equation}
\rho(\omega)
=
\begin{cases}
\displaystyle
\frac{m^{3/2}}{2\sqrt{2}\pi^2}\left[\theta(\omega-\Delta)-\theta(-\omega-\Delta)\right]\left[\sqrt{\sqrt{\omega^2-\Delta^2}+\mu}\left(\frac{\omega}{\sqrt{\omega^2-\Delta^2}}+1\right)\right.\\
\ \left.+\theta(\mu^2+\Delta^2-\omega^2)\sqrt{-\sqrt{\omega^2-\Delta^2}+\mu}\left(\frac{\omega}{\sqrt{\omega^2-\Delta^2}}-1\right)\right]\quad (\mu>0),\\
\displaystyle
\frac{m^{3/2}}{2\sqrt{2}\pi^2}\left[\theta(\omega-\sqrt{\mu^2+\Delta^2})-\theta(-\omega-\sqrt{\mu^2+\Delta^2})\right]\\
\ \times\left[\frac{\omega}{\sqrt{\omega^2-\Delta^2}}+1\right]\sqrt{\sqrt{\omega^2-\Delta^2}-|\mu|}\quad(\mu<0).
\end{cases}
\label{BCSDOS}
\end{equation}
In the BCS regime where $\mu>0$, the singularity in $\omega/\sqrt{\omega^2-\Delta^2}$ in the upper equation gives the diverging coherence peaks at $\omega=\pm\Delta$. In contrast, the singularity at $|\omega|=\Delta$ in the lower equation is less important in the BEC regime when $\mu<0$, because DOS is finite only when $|\omega|\ge\sqrt{\Delta^2+\mu^2}>\Delta$. Since the negative $\mu$ in the BEC regime is a strong-coupling effect associated with the formation of tightly bound molecules\cite{Leggett,NSR}, the suppression of the coherence peaks in the BEC regime at $T=0$ may be also regarded as a strong-coupling effect. From the above discussion, we find that the coherence peaks in DOS may be used to determine the region where strong-coupling effects are less important and one can discuss superfluid properties to some extent within the {\it weak-coupling mean-field} BCS theory.
\par

\begin{figure}
\includegraphics[width=10cm]{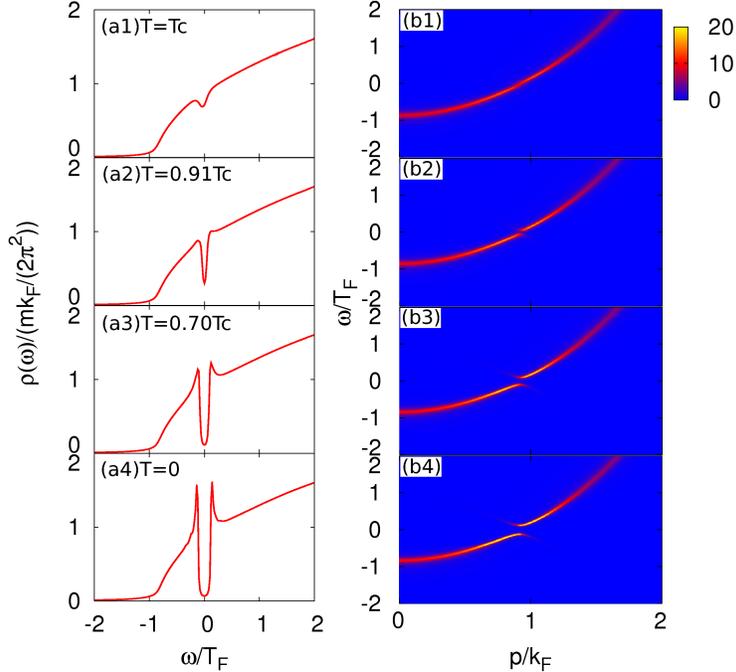}
\caption{(Color online) Temperature dependence of DOS $\rho(\omega)$
 and intensity of spectral weight $A({\bm p},\omega)$ in the weak-coupling BCS regime ($(k_Fa_s)^{-1}=-1$).
}
\label{fig5}
\end{figure}

\begin{figure}
\includegraphics[width=10cm]{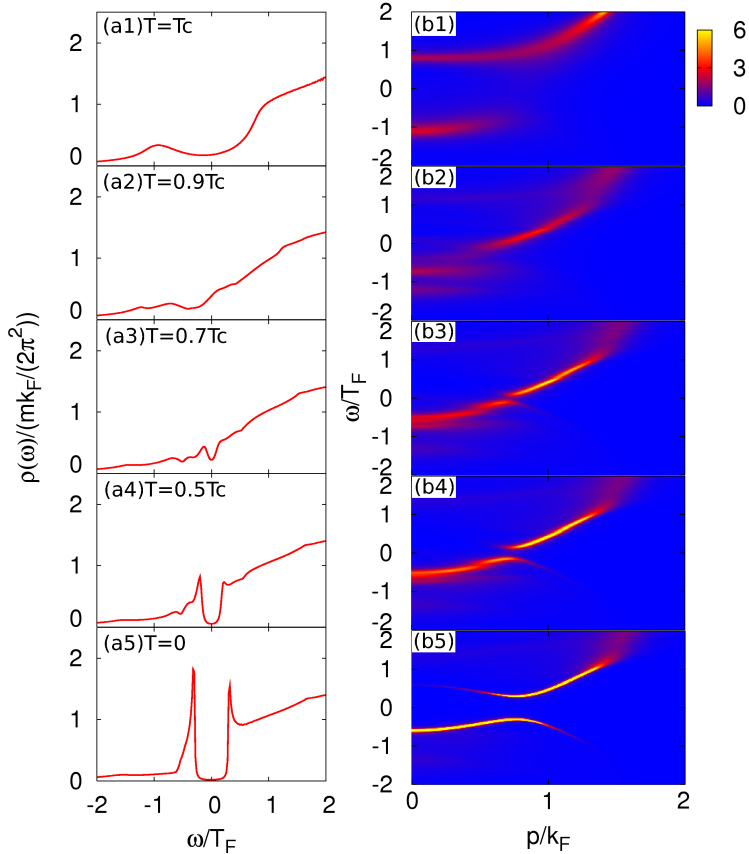}
\caption{(Color online)
Same plot as in Fig.~\ref{fig5} for $(k_Fa_s)^{-1}=0$ (unitarity limit).}
\label{fig6}
\end{figure}

\begin{figure}
\includegraphics[width=10cm]{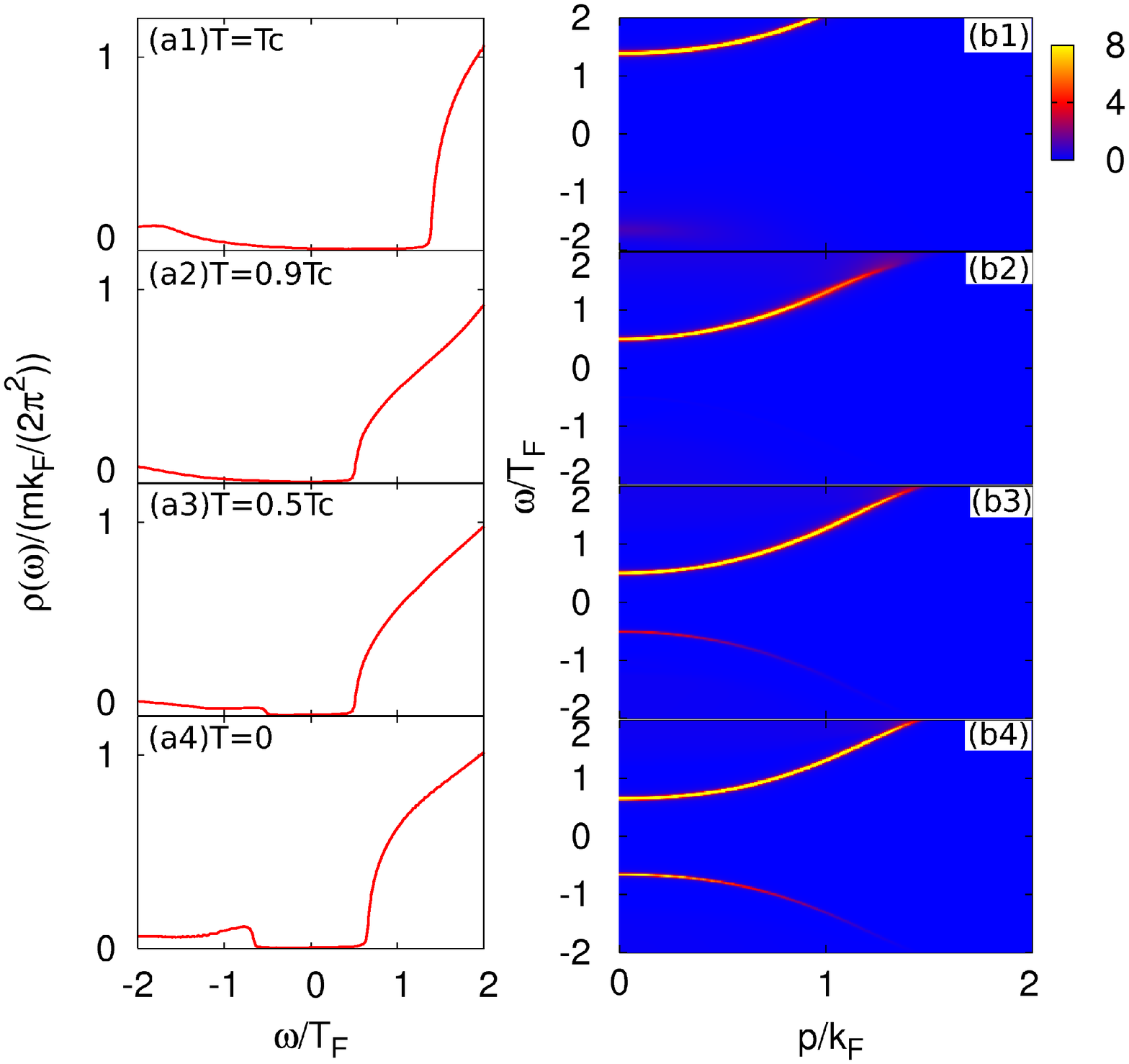}
\caption{(Color online)
Same plot as in Fig.~\ref{fig5} for $(k_Fa_s)^{-1}=0.8$ (BEC regime).}
\label{fig7}
\end{figure}
\par
We now consider the superfluid DOS $\rho(\omega)$ below $T_{\rm c}$. Figures \ref{fig5}-\ref{fig7} show the temperature dependence of calculated $\rho(\omega)$ in the BCS-BEC crossover. In the weak-coupling BCS regime (Fig.~\ref{fig5}), the pseudogap at $T_{\rm c}$ is found to smoothly change into the superfluid gap below $T_{\rm c}$. One can see the growth of the coherence peaks in panels (a2) and (a3), and the BCS-type superfluid DOS having a clear gap structure with sharp coherence peaks is eventually realized far below $T_{\rm c}$ (panel (a4)). We note that a similar continuous evolution from the pseudogap to the superconducting gap has been observed in the underdoped regime of high-$T_\mathrm{c}$ cuprates\cite{Fischer,Renner}.
\par
Figure \ref{fig5} also shows the intensity of SW in the right panels. As discussed in our previous paper\cite{Tsuchiya},  the pseudogap phenomenon in SW is not remarkable in the BCS regime. Indeed, a peak line corresponding to the free particle dispersion $\omega=p^2/2m-\mu$ is only seen in panel (b1), although the pseudogap can be clearly seen in panel (a1). The expected superfluid gap simply opens at $\omega=0$ below $T_{\rm c}$, as shown in panels (b2)-(b4). Comparing these results with the mean-field expression,
\begin{equation}
A({\bm p},\omega)=
{1 \over 2}
\left[
1+{\varepsilon_{\bm p}-\mu \over E_{\bm p}}
\right]\delta(\omega-E_{\bm p})
+
{1 \over 2}
\left[
1-{\varepsilon_{\bm p}-\mu \over E_{\bm p}}
\right]\delta(\omega+E_{\bm p}),
\label{BCSSPE}
\end{equation}
we find that the overall behavior of SW in the BCS regime is essentially the same as that in the mean-field theory.
\par
We obtain quite different results in the unitarity limit shown in Fig.~\ref{fig6}. In this case, panel (a2) clearly shows that the superfluid gap structure is still absent around $\omega=0$ even at $T=0.9T_{\rm c}$. This is because, although the superfluid order parameter $\Delta$ itself is finite, the superfluid gap structure in DOS is smeared out by strong pairing fluctuations at this temperature. However, panel (a2) also shows that pseudogap structure becomes obscure, indicating the suppression of pairing fluctuations (although they are still strong enough to smear out the superfluid gap in DOS).
\par
This suppression of pseudogap below $T_{\rm c}$ can be also seen in
SW. In Fig.~\ref{fig6}(b1), we can see the typical pseudogap structure of SW,
namely, the double-peak structure consisting of a positive energy
(particle) and negative energy (hole) branches. At $T=0.9T_{\rm c}$
(panel (b2)), this double-peak structure becomes obscure due to the
appearance of finite spectral intensity around $\omega=0$. However, the
superfluid gap still does not open at $\omega=0$ at this temperature, being consistent with panel (a2).
\par
At lower temperatures, when the pseudogap in DOS almost disappears, Fig.~\ref{fig6}(a3) shows that a dip structure appears around $\omega=0$. Correspondingly, SW also has a gap structure at $\omega=0$, as shown in panel (b3). These superfluid gap structures develop at lower temperatures, and they eventually reduce to the BCS-type DOS and SW far below $T_{\rm c}$, as shown in panels (a4) and (b4), respectively\cite{noteY}. 
\par
The above results indicate that, in the crossover region, the superfluid
gap appears in DOS after the pseudogap almost disappears. This is quite
different from the continuous evolution from the pseudogap to the superfluid gas in the weak-coupling BCS regime. Since strong pairing fluctuations, which is essential for the pseudogap phenomenon, must be suppressed to obtain the superfluid gap in DOS and SW, the evolution from the pseudogap to superfluid gap is a competing phenomenon in the crossover region. 
\par
In the strong-coupling BEC regime, since tightly bound molecules have been already formed far above $T_{\rm c}$, DOS has a clear gap structure even at $T_{\rm c}$, as shown in Fig.~\ref{fig7}(a1). Because of this clear gap structure, although the shrinkage of this gap can be seen in panel (a2) (which is considered to correspond to the suppression of the pseudogap discussed in Fig.~\ref{fig6}), one cannot precisely determine the temperature where the superfluid gap structure at $\omega=0$ starts to appear in Fig.~\ref{fig7}. In this regard, we note that the chemical potential $\mu$ is negative in the case of Fig.~\ref{fig7}. Thus, as will be discussed in Sec. IV, one should regard the system in this regime as a molecular Bose gas, rather than a Fermi gas.
\par
Figures \ref{fig7}(b3) and \ref{fig7}(b4) show the appearance of a sharp negative energy branch in SW far below $T_{\rm c}$.
This means that the overall spectral structure becomes close to the mean-field superfluid result given by Eq. (\ref{BCSSPE}) far below $T_{\rm c}$ (although the chemical potential $\mu$ remarkably deviates from the Fermi energy $\epsilon_{\rm F}$.)
For the appearance of this negative energy (hole) branch, we briefly note that it is absent in the BEC limit where the molecular formation occurs within the simple two-body physics.
On the other hand, SW has both a positive energy (particle) and negative (hole) branches in the BCS regime, reflecting that Cooper pairs are many-body bound states assisted by Fermi surface.
Thus, the sharp hole branch in panels (b3) and (b4) indicates that the many-body effect still contributes to pair formation to some extent even in the BEC regime at $(k_{\rm F}a_{\rm s})^{-1}=0.8$.
\par
In the strong-coupling BEC limit, the Green's function in Eq. (\ref{Green}) reduces to\cite{Pieri,Haussmann2}
\begin{equation}
G_{\bm p}(i\omega_n)|_{11}=\frac{-i\omega_n-\xi_p}{\omega_n^2+\xi_p^2+\Delta_{\rm
 PG}^2+\Delta^2}.
\label{GBEClim}
\end{equation}
(We summarize the derivation of Eq. (\ref{GBEClim}) in the appendix.) Here, 
\begin{equation}
\Delta=\sqrt{8\pi n_{\rm B}^0 \over m^2a_s}
\label{condbec}
\end{equation} 
is the superfluid order parameter in the BEC limit, and $\Delta_{\rm PG}=\sqrt{8\pi\tilde n_{\rm B}/m^2a_s}$ is the pseudogap parameter\cite{Pieri}, where $n_{\rm B}^0$ and ${\tilde n_{\rm B}}$ represent the molecular condensate density and molecular non-condensate density, respectively. Equation (\ref{GBEClim}) shows that the single-particle excitation gap $E_G$ is given by 
\begin{equation}
E_G=\sqrt{\mu^2+(\Delta^2+\Delta_{\rm PG}^2)}
=\sqrt{\left({1 \over 2ma_s^2}\right)^2+{4\pi n \over m^2a_s}},
\label{limitbec}
\end{equation}
where $n=2[n_{\rm B}^0+{\tilde n_{\rm B}}]$ is the total fermion density, and we have used the expression $\mu=-1/2ma_s^2$ in the BEC limit\cite{Randeria2}. Equation (\ref{limitbec}) means that the excitation gap in DOS becomes $T$-independent deep inside the BEC regime. Indeed, Fig.~\ref{fig4}(b) shows that $E_{\rm PG}\simeq E_{\rm SF}(\simeq E_{\rm G})$ when $(k_{\rm F}a_{\rm s})^{-1}\gtrsim3$.
In the extreme BEC limit ($a_s^{-1}\to+\infty$), Eq. (\ref{limitbec}) reduces to the half of the binding energy of a two-body bound state $E_{\rm bind}=1/ma_s^2$, as expected. 
\par
Before ending this section, we note that, although the overall structure
of DOS is very close to the BCS-type DOS at $T=0$ (See the lower panels
in Fig.~\ref{fig3}.), the gap size $E_{\rm SF}$ in DOS at $T=0$ is
smaller than the magnitude of superfluid order parameter evaluated in
the BCS-Leggett crossover theory\cite{Leggett} (which consists of the
mean-field gap equation and mean-field number equation), as shown in
Fig.~\ref{fig4}(a). This is because the self-energy correction $\Sigma_{\bm
p}(i\omega_n)$ in Eq. (\ref{Green}) still affects single-particle
excitations even at $T=0$, although pairing fluctuations are suppressed
far below $T_{\rm c}$. Indeed, in the present $T$-matrix theory, the
superfluid gap in DOS is affected by the off-diagonal self-energy
$\Sigma_{\bm p}(i\omega\to\omega+i\delta)|_{12}$ at $\omega\sim \Delta$
even far below $T_{\rm c}$. In addition, the present strong-coupling
theory involves effects of an effective molecular interaction within the
Born approximation, namely, the effective molecular scattering length
equals $a_{\rm B}=2a_{\rm s}$\cite{Randeria2} in the BEC regime. This
effective boson-boson interaction leads to the quantum depletion $n_d$
(which describes the number of non-condensate fermions at $T=0$), which decreases the condensate fraction $n_B^0=[n-n_d]/2$, as well as the magnitude of superfluid order parameter in the BEC regime given by Eq. (\ref{condbec}). These effects are completely ignored in the BCS-Leggett crossover theory\cite{Leggett}, so that the superfluid gap size $E_{\rm SF}$ in DOS becomes smaller than $\Delta$ evaluated in the mean-field-type crossover theory. However, since these strong-coupling effects are eventually suppressed deep inside the BEC regime, the difference between the two theories becomes small in the BEC limit, as shown in Fig~\ref{fig4}(b).
\par

\begin{figure}
\includegraphics[width=10cm]{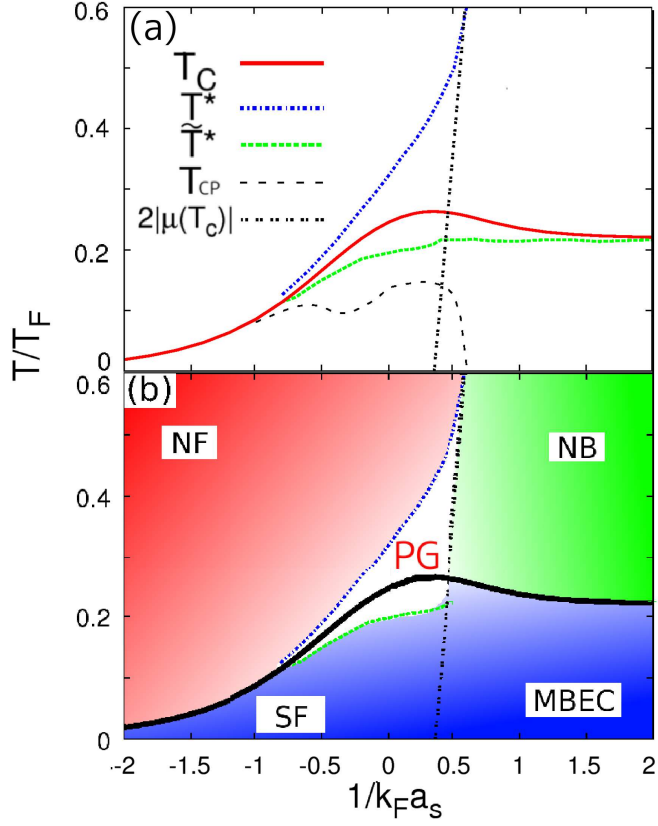}
\caption{(Color online) Phase diagram of a cold Fermi gas in the BCS-BEC crossover. (a) Characteristic temperatures introduced in this paper. ${\tilde T}$ is the temperature below which the superfluid gap appears in DOS\cite{noteZZ}. $2|\mu(T_{\rm c})|$ in the BEC regime ($\mu<0$) gives a characteristic temperature below which thermal dissociation of bound molecules is suppressed. Thus, the right side of this line may be regarded as the region of tightly bound molecular Bose gas, rather than a Fermi gas. $T^*$ is the pseudogap temperature obtained in Ref.~\cite{Tsuchiya}, where the pseudogap structure starts to appear in DOS above $T_{\rm c}$. In addition to these characteristic temperatures, we also introduce $T_{\rm cp}$ as the temperature at which the BCS-type coherence peaks appear in DOS. Below $T_{\rm cp}$, the system becomes close to the simple weak-coupling BCS state, at least with respect to single-particle excitations. (b) Phase diagram of a cold Fermi gas. PG: pseudogap phase. NF: normal Fermi gas. NB: normal state molecular Bose gas\cite{Tsuchiya}. SF: superfluid Fermi gas with a superfluid gap in DOS. MBEC: BEC of molecular bosons. We emphasize that only $T_{\rm c}$ is the phase transition temperature, and the others are all crossover temperatures without being accompanied by any phase transition. 
}
\label{fig8}
\end{figure}

\section{Phase Diagram in the BCS-BEC Crossover}
\label{section4}
Figure \ref{fig8} shows the phase diagram of a cold Fermi gas in the BCS-BEC crossover.
In panel (a), we introduce three characteristic temperatures, ${\tilde T}^*$, $T^*$, and $2|\mu(T_{\rm c})|$, in order to conveniently identity the region where pairing fluctuations dominate over single-particle properties. ${\tilde T}^*$ is the temperature where the superfluid gap appears in DOS below $T_{\rm c}$\cite{noteZZ}.
The region above ${\tilde T}^*$ is considered to be dominated by strong pairing fluctuations even in the superfluid state. $T^*$ is the so-called pseudogap temperature discussed in our previous paper\cite{Tsuchiya}, where the pseudogap starts to emerge in DOS above $T_{\rm c}$.
In addition, we also take into account the fact that physical properties in the strong-coupling BEC regime are close to those of a molecular Bose gas, rather than a Fermi gas. Noting that the molecular binding energy $E_{\rm bind}$ in this regime is deeply related to the Fermi chemical potential as $E_{\rm bind}\simeq 2|\mu|$ when $\mu<0$, one may expect that the thermal dissociation of molecules is suppressed in the BEC regime when $T\lesssim 2|\mu|$ ($\mu<0$). Thus, it is convenient to regard the right side of the $2|\mu(T_{\rm c})|$-line in Fig.~\ref{fig8}(a) as the molecular Bose gas regime\cite{Tsuchiya}.
We briefly note that $\tilde{T}^*$, $T^*$, and $2|\mu(T_{\rm c})|$, are all crossover temperatures without being accompanied by any phase transition.
\par
Using these three characteristic temperatures in Fig.~\ref{fig8}(a), we find that the region ``PG''  in Fig.~\ref{fig8}(b), which is surrounded by $T^*$, ${\tilde T}^*$, and $2|\mu(T_{\rm c})|$, is the one where the pseudogap structure in DOS is remarkable. Thus, we conveniently call this region the {\it pseudogap region} (although, strictly speaking, the region below $T_{\rm c}$ is the superfluid state). In this pseudogap regime, strong pairing fluctuations induce a gap-like structure in DOS in both the normal and superfluid phases.
\par
Below ${\tilde T}^*$ (``SF'' in Fig.~\ref{fig8}(b)), instead of the disappearance of the pseudogap, the superfluid gap starts to develop in DOS, so that single-particle properties are dominated by superfluid gap.
As one further decreases the temperature below ${\tilde T}^*$, one eventually obtains the weak-coupling BCS-type DOS characterized by a clear excitation gap and coherence peaks, as discussed in Sec. III.
To conveniently include this, we also introduce the characteristic temperature $T_{\rm cp}$ at which the coherence peaks appear in DOS\cite{noteZZZ} in Fig.~\ref{fig8}(a).
Below $T_{\rm cp}$, single-particle properties are close to those in the weak-coupling mean-field BCS state.
\par
In the molecular Bose gas regime (``NB'' and ``MBEC'' in Fig.~\ref{fig8}(b)), a large single-particle excitation gap already exists above $T_{\rm c}$, reflecting a large molecular binding energy ($E_{\rm bind}\simeq 2|\mu|\simeq 1/ma_s^2$\cite{Randeria2}). This large binding energy suppresses single-particle excitations accompanied by pair breaking in the superfluid phase below $T_{\rm c}$, so that excitations are dominated by collective Bogoliubov modes, as in the case of Bose superfluid. In this sense, we call the superfluid region in the molecular Bose gas regime the BEC of molecular bosons (``MBEC'' in Fig \ref{fig8}(b)).
\par

\section{summary}

To summarize, we have investigated single-particle excitations and
strong-coupling effects in the BCS-BEC crossover regime of a superfluid
Fermi gas. Extending our previous work above $T_{\rm c}$ to the
superfluid phase below $T_{\rm c}$, we have numerically calculated the
superfluid DOS, as well as SW,
within the $T$-matrix theory. We have systematically examined
how the pseudogap at $T_{\rm c}$ evolves into the superfluid gap, as one
decreases the temperature below $T_{\rm c}$. While the evolution is
continuous in the weak-coupling BCS regime, the superfluid gap was shown
to appear in DOS after the pseudogap is suppressed below $T_{\rm c}$ in
the crossover regime. Using these results, we have identified the
pseudogap region where strong pairing fluctuations dominate over
single-particle properties in the phase diagram of a cold Fermi gas.
\par
Since the observation of single-particle excitations has recently become
possible in cold Fermi gases by photoemission-type experiment,
measurements of single-particle excitation spectrum affected by strong
pairing fluctuations discussed in this paper would be an interesting
problem to understand the strong-coupling superfluid properties in the
BCS-BEC crossover.
\par
In this paper, we have assumed a uniform Fermi gas, for simplicity. In a
trapped system, it is an interesting problem how the spatial
inhomogeneity affects the evolution from the pseudogap to the superfluid
gap below $T_{\rm c}$. Since a real Fermi gas is always trapped in a
harmonic potential, this is also an important issue in comparing
experimental data with theoretical calculations. We will
discuss this problem in a future paper.

\acknowledgments
We thank Y. Yanase, S. Watabe, D. Inotani and T. Kashimura for fruitful discussions. This work was supported by the Japan Society for the Promotion of Science, and Global COE Program ``High-Level Global Cooperation for Leading-Edge Platform on Access Spaces (C12)''.

\appendix
\section{Analytic results in strong-coupling BEC limit}

In this appendix, we present the outline of the derivation of Eq. (\ref{GBEClim}). For more details, we refer to Refs.~\cite{Pieri,Haussmann2}. In the BEC limit, the particle-particle scattering matrix in Eq. (\ref{gamma}) reduces to\cite{Pieri,Haussmann2}
\begin{eqnarray}
\left(
\begin{array}{cc}
\Gamma_{\bm q}^{+-}(i\nu_n)&
\Gamma_{\bm q}^{++}(i\nu_n)\\
\Gamma_{\bm q}^{--}(i\nu_n)&
\Gamma_{\bm q}^{-+}(i\nu_n)\\
\end{array}
\right)\simeq
{8\pi \over m^2a_s}
{1 \over \nu_n^2+{E_{\bm q}^{\rm B}}^2}
\left(
\begin{array}{cc}
-i\nu_n+q^2/4m-\mu_B&
-\mu_B
\\
-\mu_B&
i\nu_n+q^2/4m-\mu_B
\end{array}
\right),
\nonumber
\\
\label{gammapmBEC}
\end{eqnarray}
where $E_{\bm q}^{\rm B}=\sqrt{\frac{q^2}{4m}(\frac{q^2}{4m}+2|\mu_{\rm B}|)}$
is the Bogoliubov excitation spectrum of a molecular BEC, and $\mu_{\rm B}=-\Delta^2/4|\mu|$ is the Bose chemical potential. The number equation in the BEC limit reduces to $N=2(n_{\rm B}^0+\tilde n_{\rm B})$, where $n_{\rm B}^0$ and $\tilde n_{\rm B}$ are the molecular condensate and non-condensate density, respectively, given by
\begin{eqnarray}
n_{\rm B}^0&=&{1 \over \beta}\sum_{\bm q,\nu_n}\frac{\mu_{\rm
 B}}{\nu_n^2+{E_{\bm q}^{\rm B}}^2},\\
\tilde n_{\rm B}&=&{1 \over \beta}\sum_{\bm q,\nu_n}\frac{i\nu_n+q^2/4m-\mu_{\rm B}}{\nu_n^2+{E_{\bm q}^{\rm B}}^2}.
\end{eqnarray}
Using Eq.~(\ref{gammapmBEC}), we approximate the self-energy in Eq.~(\ref{self-energy}) to
\begin{eqnarray}
\Sigma_{\bm p}(i\omega_n)
\simeq
{8\pi \over m^2a_{\rm s}}
\left(
\begin{array}{cc}
{\tilde n}_B G^0_{\bm p}(i\omega_n)|_{22}& 
n_B^0 G^0_{\bm p}(i\omega_n)|_{12}\\
n_B^0 G^0_{\bm p}(i\omega_n)|_{21}&
{\tilde n}_B G^0_{\bm p}(i\omega_n)|_{11}\\
\end{array}
\right),
\label{selfeBEC2}
\end{eqnarray}
where we have approximately set ${\bm q}=\nu_n=0$ in $G_{\bm p+\bm q}^0(i\omega_n+i\nu_n)$. Using Eq. (\ref{selfeBEC2}), one obtains the diagonal component of the Green's function as
\begin{eqnarray}
G_{\bm p}(i\omega_n)|_{11}=
{1 
\over 
\displaystyle
i\omega_n-\xi_p-{8\pi{\tilde n}_B \over m^2a_{\rm s}}
G_{\bm p}^0(i\omega_n)|_{22}-
{
\displaystyle
\left[
\Delta-{8\pi n_{\rm B}^0 \over m^2a_{\rm s}}
G_{\bm p}^0(i\omega_n)|_{12}
\right]^2
\over 
\displaystyle
i\omega_n+\xi_p-{8\pi{\tilde n}_{\rm B} \over m^2a_{\rm s}}
G_{\bm p}^0(i\omega_n)|_{11}
}
}.
\label{gbec}
\end{eqnarray}
Expanding the denominator in Eq. (\ref{gbec}) up to $O(\Delta^2)$, we obtain Eq. (\ref{GBEClim}).



\begin{thebibliography}{99}
\bibitem{noteZ} Since $\rho_1({\bm q})$ describes amplitude fluctuations
	of the order parameter around the mean-field value, we actually
	subtract $\langle\rho_1({\bm q}=0)\rangle$ from $\rho_1({\bm q})$ (although we do not explicitly show this in the text).
\bibitem{noteY} We briefly note that, although the upper branch in SW at $T=0$ looks to be smeared out at $\omega \gtrsim 1$, this peak line actually becomes sharp and narrow at higher energy $\omega\gtrsim 2$.
\bibitem{noteZZ} The characteristic temperature ${\tilde T}^*$ becomes
	difficult to be determined in the BEC regime, because a large excitation gap already exists at $T_{\rm c}$ in this regime. Because of this problem, we only show ${\tilde T}^*$ in the region $(k_{\rm F}a_s)^{-1}\le 0.5$.
\bibitem{noteZZZ} We have determined $T_{\rm cp}$ as the temperature where the upper coherence peak appears in DOS.
\end{thebibliography}
\end{document}